# pPython Performance Study

Chansup Byun, William Arcand, David Bestor, Bill Bergeron, Vijay Gadepally, Michael Houle, Matthew Hubbell,
Hayden Jananthan, Michael Jones, Anna Klein, Peter Michaleas, Lauren Milechin, Guillermo Morales,
Julie Mullen, Andrew Prout, Albert Reuther, Antonio Rosa, Siddharth Samsi, Charles Yee, Jeremy Kepner
Massachusetts Institute of Technology

*Abstract*—pPython seeks to provide a parallel capability that provides good speed-up without sacrificing the ease of programming in Python by implementing partitioned global array semantics (PGAS) on top of a simple file-based messaging library (PythonMPI) in pure Python. pPython follows a SPMD (single program multiple data) model of computation. pPython runs on a single-node (e.g., a laptop) running Windows, Linux, or MacOS operating systems or on any combination of heterogeneous systems that support Python, including on a cluster through a Slurm scheduler interface so that pPython can be executed in a massively parallel computing environment. It is interesting to see what performance pPython can achieve compared to the traditional socket-based MPI communication because of its unique file-based messaging implementation. In this paper, we present the point-to-point and collective communication performances of pPython and compare them with those obtained by using mpi4py with OpenMPI. For large messages, pPython demonstrates comparable performance as compared to mpi4py.

I. INTRODUCTION

Python is one of the most widely used programming languages among developers around the world [1]. Python has become popular among scientific and engineering computing communities because Python is open source, its syntax is easy to understand, and it has a rich ecosystem of scientific and mathematical packages, such as NumPy [2] and SciPy [3]. Furthermore, many different approaches have been developed for parallel programming with Python on shared and distributed memory environments. A comprehensive list of parallel Python libraries and software is available in Ref. 4.

Among the various approaches cited in Ref. 4, the message passing approach [5-10] appears to be one of the most widely used approaches. The message passing approach requires the user to explicitly send messages within the code. These approaches often implement a variant of the Message Passing Interface (MPI) standard [11]. Message passing allows any processor to directly communicate with any other processor and provides the minimum required functionality to implement a parallel program. Users that are already familiar with MPI find these approaches powerful. However, the learning curve is steep for the typical user because explicit message passing approaches significantly lower the level of abstraction and require users to deal directly with deadlocks, synchronization, and other low level parallel programming concepts. In addition, the impact on code size is significant. Serial programs converted to parallel programs with MPI typically increase in size by 25% to 50%; in contrast, OpenMP and PGAS approaches typically increase the code size by only ~5% [12].

In spite of these difficulties, a message passing capability is a requirement for other parallel programming approaches such as client-server or global arrays for distributed memory programming. Furthermore, message passing is often the most efficient way to implement a program and there are certain programs with complex communication patterns that can only be implemented with direct message passing. Therefore, any complete parallel solution must provide a mechanism for accessing the underlying messaging layer. Among the available Python message passing implementations, mpi4py [5] is widely used [13] and is actively maintained.

Although a number of other parallelization approaches have been developed for Python [4], there are a limited number of published works on the parallelization of Python with global arrays. The extensive list of work related to global arrays is surveyed in Ref. 14. In particular, there have been a couple of publications on implementing global arrays in Python [13,15]. Recently, we have published a PGAS implementation called pPython [29] to provide a parallel capability with good speed-up without sacrificing the ease of programming in Python.

The MIT Lincoln Laboratory Supercomputing Center (LLSC) has focused on developing a unique, interactive, on-demand high-performance computing (HPC) environment to support diverse science and engineering applications. This system architecture has evolved into the MIT SuperCloud. MIT SuperCloud not only continues to support parallel MATLAB and Octave jobs, but also jobs in Python [16], Julia [17], R [18], TensorFlow [19], PyTorch [20], and Caffe [21] along with parallel C, C++, Fortran, and Java applications with various flavors of message passing interface (MPI) [11].

One of the core software stacks at LLSC environment is pMatlab [22,23], which implements Partitioned Global Array Semantics (PGAS) [24] using standard operator overloading techniques. pMatlab includes MatlabMPI [25], which provides MPI point-to-point communication and the gridMatlab [26] scheduler interface. pMatlab has subsequently inspired the MathWorks parallel computing toolbox used by many thousands of scientists and engineers around the world. pPython seeks to provide all the benefits available with pMatlab, MatlabMPI, and gridMatlab in a Python programming environment.

The core data structure in pPython is a distributed numerical array whose distribution onto multiple processors is specified with a 'map' construct. Communication operations between distributed arrays are abstracted away from the user and pPython transparently supports redistribution between any block-cyclic-overlapped distributions in up to four dimensions. pPython is built on top of the PythonMPI communication library and runs on any combination of heterogeneous systems that support Python, which includes Windows, Linux, and MacOS operating systems. In addition, pPython includes a scheduler interface that enables users to submit their computing tasks via the scheduler.

In the past, we have compared pPython's performance against that of pMatlab's (which is based on similar technology) using the HPC Challenge parallel computing benchmarks STREAM, FFT, High Performance Linpack (HPL), and RandomAccess. The results are well matched between the two different language implementations. [29]. As an extension of the pPython performance study, we have conducted some additional performance testing with basic MPI communications such as point-to-point and collective communications such as broadcast and aggregation. In pPython, PythonMPI library is used to enable message communications among MPI processes PythonMPI implements only a small set of MPI functions such as MPI_Send and MPI_Recv for point-to-point communication and MPI_Bcast for message broadcasting to all MPI processes. In addition, an aggregation function called agg() is implemented to aggregate a distributed array to the leader MPI process. In this paper, we have studied the performance of those basic MPI functions available in pPython and compared the results with those obtained by using the mpi4py package, which implements the MPI standard in Python with an underlying MPI library available on the system.

## II. pPython Interface and Architecture Design

pPython interface and architecture design are implemented in pure Python and provide ease-of-use, high performance, and ease-of-implementation. Although the details of each of these features are well discussed in the earlier work [29], a brief summary is described below.

### A. Ease-of-Use

pPython adopts a separation-of-concerns approach to make program correctness and mapping a program to a parallel architecture orthogonal. A serial program is made parallel by adding maps to arrays. Maps only contain information about how an array is broken up onto multiple processors and the addition of a map does not change the functional correctness of a program. A map (Fig. 1) is composed of a grid specifying how each dimension is partitioned, a distribution that selects either a block, cyclic or block-cyclic partitioning, and a list of processor IDs that defines which processors actually hold the data.

In Python, when an array is created, its default ordering is row-major order (C-style). In order to provide flexibility with ordering the processor grid, the keyword `order` is introduced with the map function. This allows it to support a processor grid arranged in the column–major order (Fortran style). It is also noted that the map function name itself is also changed to 'Dmap' because Python already has a 'map' function for a different purpose. In pPython, when creating a distributed array as shown in Fig. 1, the keyword `map` is introduced for a distributed array. Without the keyword `map`, the distributed array functions, `zeros`, `ones`, and `rand`, will return as standard NumPy arrays.

PGAS enables complex data movements to be expressed compactly without making parallelism a burden to code. For example, removing the maps from pPython code returns the program to a valid serial program that simply uses standard built-in operations. This is a direct result of the orthogonality of mapping and functionality, and allows the pPython library to be "turned off" by simply setting all the maps equal to the scalar value of one.

### B. High Performance

pPython adopts a coding style that uses fragmented PGAS constructs (see Ref. 22 for definition of fragmented PGAS and details about the choice). This style is less elegant but provides strict guarantees on performance. More specifically, distributed arrays are used as little as possible and only when interprocessor communication is required.

However, PGAS constructs are not appropriate for all circumstances. There are communication patterns that would be more efficient if direct message passing could be employed. Thus, it is important to have mechanisms that allow PGAS and the underlying communication constructs to interact easily. pPython provides this feature by allowing the user to directly access the underlying PythonMPI library and its data structures. Several of the HPC Challenge benchmarks fall into the class of codes that do best by allowing some use of direct message passing [22]. HPC Challenge parallel computing benchmarks STREAM, FFT, High Performance Linpack (HPL), and RandomAccess are implemented with pPython to demonstrate this ability.

### C. Ease-of-Implementation

Considering the need to achieve balance between ease-of-use and high-performance [22], one of the key choices in implementing a PGAS library is deciding which data distributions to support. At one extreme, it can be argued that most users are satisfied by 1-D block distributions. At the other extreme, one can find applications that require truly arbitrary distributions of array indices to processors. pPython has chosen to support up to 4-D block-cyclic distributions with overlap because the problem of redistribution between any two such distributions is highly complex to program for the use but has been solved a number of times by different parallel computing technologies. This allows pPython users to create four-dimensional arrays with all four dimensions distributed.

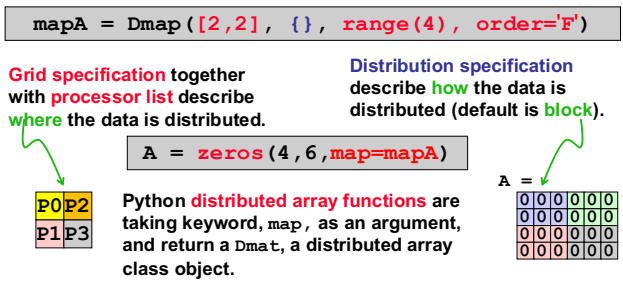

Fig. 1. Anatomy of a map. A map for a numerical array is an assignment of blocks of data to processing elements. It consists of a grid specification (in this case, {} implies that the default block distribution should be used), and a processor list (in this case the array is mapped to processors, 0., 1, 2, and 3). In pPython, grid specification can choose the ordering of processors in column or row direction with the keyword, order.

## III. pPython Implementation

This section discusses the implementation of the pPython library. pPython employs a layered architecture [22]. Additionally, pPython has added a scheduler interface for a grid environment such as the LLSC environment. In the layered architecture, the pPython library implements distributed constructs such as distributed matrices and higher dimensional arrays. pPython also provides parallel implementations of a select number of functions such as redistribution and Fast Fourier Transform (FFT).

The pPython library uses the parallelism through polymorphism approach as discussed by Choy and Edelman [27]. In addition, the polymorphism is further exploited by introducing the map class object. Map objects belong to a pPython `Dmap` class and are created by specifying the grid description, distribution description, and the processor list (Fig. 1). The map object can then be passed to a pPython method, such as `rand`, `zeros`, or `ones`. These methods are implemented in a way that, when a map object is passed, the library creates a distributed array class, `Dmat`, object.

Since all functions supported in pPython are implemented in pure Python, pPython can run anywhere Python runs, given that there exists a common file system, a constraint imposed on pPython by the PythonMPI file-based messaging library. There are some benefits with the file-based communications. It can handle messages with very large sizes without any issues as long as the underlying filesystem has the disk space. It is also implemented as one-sided sends and eliminates many race conditions, which are much easier for users. In addition, security of messages is taken care of the underlying filesystem. A further benefit of the layered architecture of pPython is that any other communication library could be substituted for PythonMPI. Further details about pPython implementation have been discussed in Ref. [29].

## IV. PERFORMANCE EXPERIMENT

In this performance study, we have set aside 16 compute nodes from the TX-Green system [31] at the Lincoln Laboratory. Each compute node has two-socket 24-core Xeon Platinum 8269 at 2.40 GHz, 192 GB RAM memory, and 4 TB local disk with 25 GigE network. The Lustre parallel filesystem is running on a 14 petabyte DDN 14000 storage array for users' home directory, which is directly connected to the core switch.

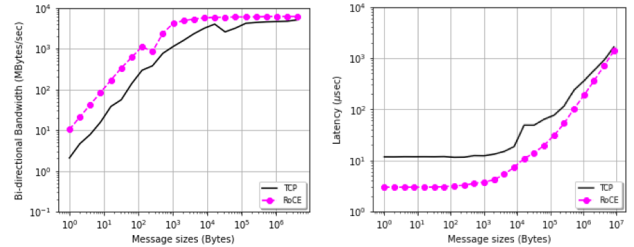

Fig. 2. MPI bandwidth and latency performance comparison when using the TCP (Transmission Control Protocol) and RoCE (RDMA over Converged Ethernet) protocol for the point-to-point message communication with sizes ranging from 1 byte to 8 Mbytes. The osu benchmark is used along with OpenMPI 4.1.3 and 25 GigE network.

In order to measure and compare pPython communication performance using the file-based messaging library, we have chosen the mpi4py python package, which is widely used by Python community for parallel Python programming. The performance metrics we have chosen for this study are a couple of basic MPI communication operations, point-to-point and collective communications. For collective communication, both aggregation and broadcasting performances are used.

Since mpi4py needs a MPI library, we have used the OpenMPI version 4.1.3 library, which is built for the RDMA over converged Ethernet (RoCE) protocol for better performance as compared to Transmission Control Protocol (TCP). The MPI communication performance comparison between the two different protocols (TCP and RoCE) has been performed using the point-to-point MPI communication benchmark from OSU Micro-Benchmarks [30]. The results are obtained by running two MPI processes, one process per node and are compared in Fig. 2. As shown in Fig. 2, the RoCE protocol outperforms the TCP protocol in bandwidth and latency performance throughout the message sizes ranging from 1 byte to 8 Mbytes. Thus RoCE protocol is used when evaluating the communication performance for mpi4py package.

### A. Point-to-point Communication

In order to study the performance of point-to-point communication, we have selected a couple of scenarios: communication between the two MPI processes co-located on the same node and between two processes, one process per each node. Also, we have used two different filesystems: the Lustre parallel filesystem and the local filesystem as a medium for the file-based message communication when studying pPython communication performance. The message sizes range from 16 bytes to 1 Gbytes.

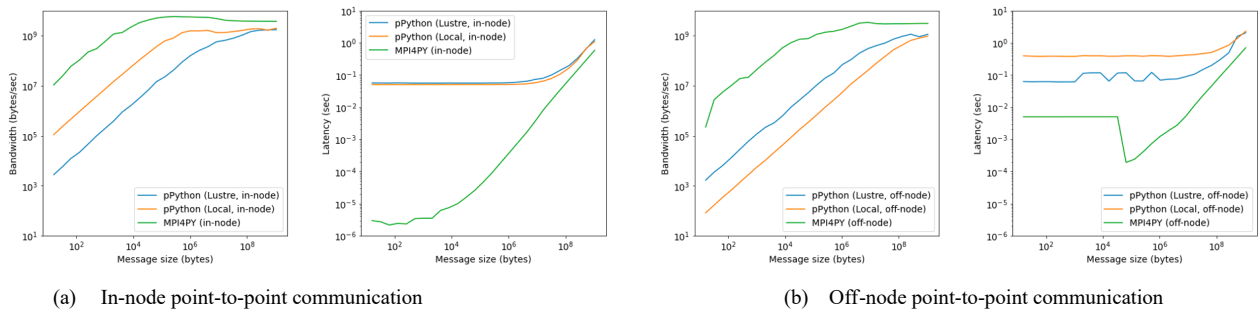

(a) In-node point-to-point communication

(b) Off-node point-to-point communication

Fig. 3. MPI bandwidth and latency performance comparison between pPython and mpi4py for the point-to-point message communication with sizes ranging from 16 byte to 1 Gbytes. The pPython results are obtained by using the Lustre parallel filesystem and a Linux local filesystem.

The results are presented in Fig. 3. We have measured the performance data 5 times for each message size and used the geometric means in the plot. Due to the overhead associated with the file IO with the file-based message communication, pPython performance is significantly lower, especially with smaller messages, than that of mpi4py across the range of message sizes we have studied in Fig. 3. In addition, OpenMPI shared memory optimization of message communication between the two processes on the same node should have contributed to the excellent performance of mpi4py as shown in case (a) in Fig. 3. It is interesting to note that the local filesystem performs better than the Lustre filesystem with pPython when the two processes are on the same node as shown in case (a). This is due to the fact that, with the Lustre filesystem, the message file is actually stored on a storage system (DDN 14K), which requires additional time in data transition over network with write and read operations as compared to same operations done on the local filesystem. Furthermore, when the message size is big enough, such as 10 Kbytes or larger, the performance gap between pPython and mpi4py becomes significantly narrower as shown in Fig. 3.

It should be noted that, in pPython triples mode [3], all the MPI processes launched on the same node is managed by dynamically-generated execution script with the process pinning. In the above in-node case in Fig. 3 (a), since there are two sockets on each machine, one MPI process is placed on socket 0 and the other MPI process on socket 1.

However, when point-to-point communication is performed between the two processes that are not on the same node as shown in Fig. 3 (b), pPython performance shows that the Lustre filesystem does better than the local filesystem. With the local filesystem, a scp command is used to send the message file across the network. This scp command introduces a large overhead for the point-to-point communication. With the Lustre filesystem, there is little change in the pPython point-to-point communication performance since the message communication overhead remains the same in both cases. It is important to note that, since the security for transferring message files is entirely handled by the scp tool and the file system permissions, no additional security or ports are required other than those that are typically required on an HPC system [32]. In this case, the mpi4py performance has dropped slightly lower than the performance observed in Fig. 3 (a) but not as much as what has been observed with pPython performance drop with the local filesystem-based messaging kernel. Based on this trend, in order to obtain the best performance with pPython, it is desirable to use the local filesystem when both processes are on the same node and the Lustre filesystem when both processes are not on the same node. Also, it is noted that there is sudden change in the latency characteristics with mpi4py point-to-point communication in Fig. 3 (b). This is a typical behavior with OpenMPI RDMA implementation where it changes its communication protocol beyond 12 Kbyte message size. [28]

*B. Collective Communication*

In pPython, we have implemented two commonly used MPI collective communications: aggregation of data on a distributed array among the MPI processes to a single MPI process and broadcasting of data from a single MPI process to the rest of the MPI processes. We have compared the performance of these functions with those obtained by using equivalent functions in mpi4py.

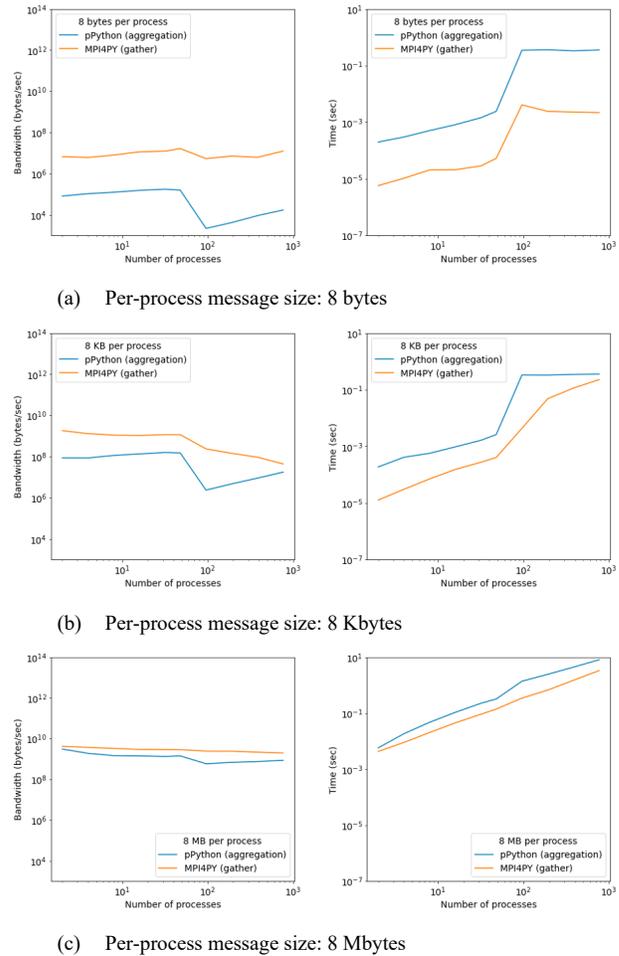

(a) Per-process message size: 8 bytes

(b) Per-process message size: 8 Kbytes

(c) Per-process message size: 8 Mbytes

Fig. 5. Aggregation performance comparison between pPython and mpi4py in terms of globl bandwidth and time with three different message sizes per process: 8 bytes, 8 Kbytes and 8 Mbytes, with various number of MPI processes ranging from 2 to 768. The pPython results are obtained by using the local filesystem-based message communication.

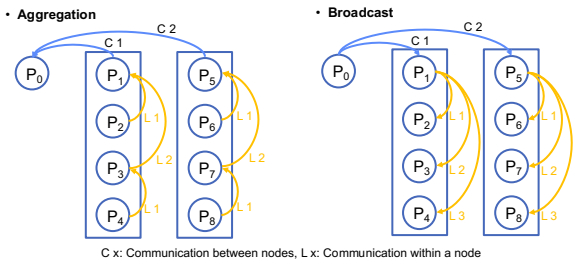

Fig. 4. pPython implementation of aggregation and broadcast operations. The implementation is based on a node-aware algorithm and the operation is separated into two levels: in-node operation and across-node operation. The initial implementation used a binary-based aggregation but then used serialized broadcasting at each stage of the operations.

In an initial pPython implementation, both aggregation and broadcasting operations are optimized to use a node-aware algorithm with the triples mode jobs [31] where the operations are separated into two levels of communications: in-node and off-node communications as shown in Fig. 4. Furthermore, aggregation operations between in-node and off-node processes, is progressed in the order of the MPI processes determined in a binary-tree based communication algorithm so that the aggregation process can introduce concurrency and in turn improve the performance. However, for the broadcast operation, both in-node and off-node broadcast operations are done serially as shown in Fig. 4.

*1) Aggregation*

In pPython, the agg() function aggregates a distributed array across all MPI processes onto the leader MPI process. A similar operation can be done using the MPI_gather() function available in mpi4py. In order to study the aggregation performance of pPython and compare its performance with the mpi4py gather() function, we have selected three different message sizes to represent small (8 bytes), medium (8 Kbytes) and large (8 Mbytes) messages. mpi4py outperforms pPython in aggregation performance in all three different message sizes as shown in Fig. 5. However, as shown in Fig. 5 (c), the gap between the two has narrowed significantly. For the medium and large message sizes, the pPython performance increases while the mpi4py performance decreases as the number of MPI processes scales beyond a node boundary in the aggregation. In this benchmark setup, the node boundary is 48 MPI processes. It would be interesting to see what would happen if the aggregation is performed on a scale with a much larger number of MPI processes.

It should be noted that, from the result of pPython aggregation study, there is a bandwidth performance drop as soon as the aggregation happens beyond a node boundary for all three message sizes, where the change is more severe with small and medium messages. Furthermore, for both pPython and mpi4py cases, the total time remains relatively the same for the small message size when the aggregation scales beyond a node boundary, which results in increasing the global bandwidth as the number of MPI processes increases. But, for large message sizes, the total time increases linearly as the number of MPI processes increases, and in turn, this reduces the global bandwidth performance. We believe this behavior is caused by the fact that the message travel time through the network is more dominant for small and medium messages when aggregation happens beyond a node boundary.

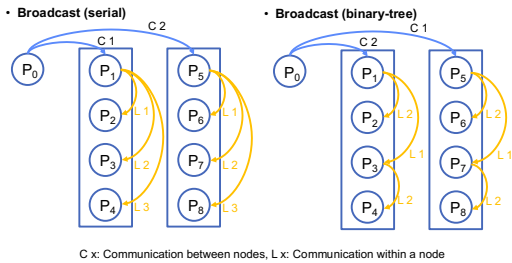

Fig. 6. pPython optimization of broadcast operation using a node-aware, binary-tree based broadcast at each stage of the two level operations, one among the leader processes of nodes and the other among the processes within each node.

*2) Broadcast*

While we are comparing pPython broadcast performance with that of mpi4py, we have found that the initial implementation of broadcast operation is not efficient because both in-node and off-node broadcast operations are serialized. In order to remove this performance bottleneck, the broadcast operation is also updated with using a node-aware, binary-tree based communication strategy as shown in Fig. 6. By implementing this communication strategy, the broadcast operation can be accomplished a lot faster in pPython as demonstrated in Fig. 7.

The node-aware, binary-tree based broadcast strategy in pPython shows comparable performance with mpi4py when the message is large (8 Mbytes in size in the experiment) and is broadcasted to the MPI processes within the same node. Also, the optimized broadcast implementation has improved its performance significantly as compared to the initial implementation in pPython for all three message sizes across the entire range of MPI processes studied in this experiment. However, significant performance drop has been observed when the broadcast operation is performed across more than a node boundary. This is mainly caused by the overhead associated with remote copy of the message using scp. Although mpi4py broadcast operation can benefit from RoCE protocol being used by the OpenMPI library under the hood, scp operation is based on TCP and is slower than RoCE protocol.

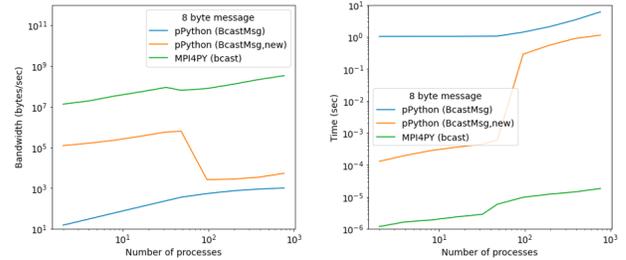

(a) Per-process message size: 8 bytes

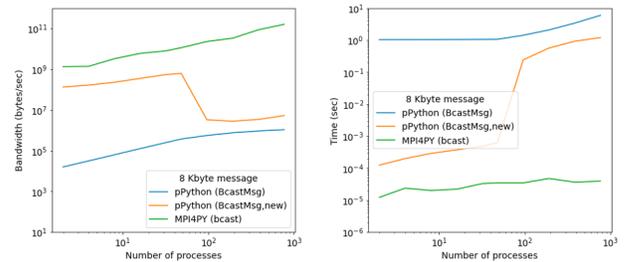

(b) Per-process message size: 8 Kbytes

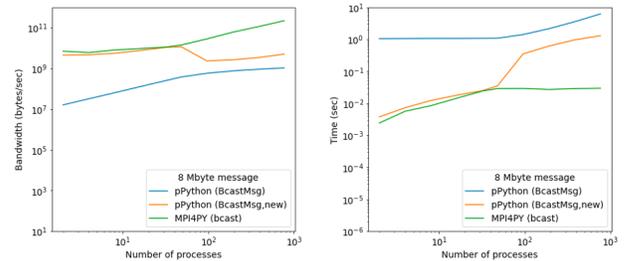

(c) Per-process message size: 8 Mbytes

Fig. 7. Broadcast performance comparison between pPython and mpi4py in terms of globl bandwidth and time with three different message sizes per process: 8 bytes, 8 Kbytes and 8 Mbytes, for various number of MPI processes ranging from 2 to 768. The pPython results were obtained with the initial and optimized implementations by using the local filesystem as a medium for message communications.

V. SUMMARY

Because of pPython's unique approach with file-based message communication, it is of interest to see what communication performance pPython can achieve and how it compares with traditional socket-based MPI communication. In this paper, we have executed pPython performance study on the MPI point-to-point and collective communications, which supplements the earlier performance study [29]. Furthermore, we have compared the results with those obtained by using MPI4PY which is a Python wrapper for MPI API (Application Programming Interface) and can be used any MPI library. In this paper, we have obtained MPI4PY results using OpenMPI 4.1.3 library which is specially built for RoCE protocol. In general, the RoCE protocol can provide better MPI communication performance with lower latency and high bandwidth as compared to TCP protocol.

Although pPython is slower than mpi4py in MPI communications because pPython uses a file-based message communication while mpi4py uses a socket-based message communication, we have demonstrated that pPython's performance can be comparable with mpi4py under certain conditions for both point-to-point and collective communications. We have further optimized the pPython broadcast communication by introducing a node-aware, binary tree-based communication algorithm, which enables comparable broadcast performance when all MPI processes are on the same node. Finally, as a future task, we have identified potential performance improvement opportunities with pPython broadcast communication by utilizing Lustre parallel filesystem for off-node process communication and local filesystem for in-node process communication.

Overall, pPython provides a simple and maintainable way to easily make Python programs run in parallel with reasonable performance scalability and comparable performance as compared to the traditional MPI communication available with mpi4py.

ACKNOWLEDGMENTS

The authors wish to acknowledge the following individuals for their contributions and support: Bob Bond, Alan Edelman, Jeff Gottschalk, Charles Leiserson, Joseph McDonald, Heidi Perry, Steve Rejto, Matthew Weiss, and Marc Zissman.